\documentclass[journal]{vgtc}                
\ifpdf
  \pdfoutput=1\relax                   
  \usepackage{graphicx}                
  \DeclareGraphicsExtensions{.pdf,.png,.jpg,.jpeg} 
\else
  \usepackage{graphicx}                
  \DeclareGraphicsExtensions{.pdf}     
\fi%

\graphicspath{{figures/}{pictures/}{images/}{./}} 

\usepackage{microtype}                 
\PassOptionsToPackage{warn}{textcomp}  
\usepackage{textcomp}                  
\usepackage{mathptmx}                  
\usepackage{times}                     
\usepackage{cite}                      
\usepackage{tabu}                      
\usepackage{booktabs}                  

\onlineid{1010}

\vgtccategory{Research}
\vgtcpapertype{technique}

\title{FoVolNet: Fast Volume Rendering using \\Foveated Deep Neural Networks}

\author{David Bauer, Qi Wu, and Kwan-Liu Ma}
\authorfooter{
\item
 David Bauer, Qi Wu, and Kwan-Liu Ma are with the University of California at Davis. E-mail: {davbauer,qadwu,klma}@ucdavis.edu.
}

\shortauthortitle{Bauer \MakeLowercase{\textit{et al.}}: FoVolNet}

\abstract{Volume data is found in many important scientific and engineering applications. Rendering this data for visualization at high quality and interactive rates for demanding applications such as virtual reality is still not easily achievable even using professional-grade hardware. We introduce FoVolNet---a method to significantly increase the performance of volume data visualization. We develop a cost-effective foveated rendering pipeline that sparsely samples a volume around a focal point and reconstructs the full-frame using a deep neural network. Foveated rendering is a technique that prioritizes rendering computations around the user's focal point. This approach leverages properties of the human visual system, thereby saving computational resources when rendering data in the periphery of the user's field of vision. Our reconstruction network combines direct and kernel prediction methods to produce fast, stable, and perceptually convincing output. With a slim design and the use of quantization, our method outperforms state-of-the-art neural reconstruction techniques in both end-to-end frame times and visual quality. We conduct extensive evaluations of the system's rendering performance, inference speed, and perceptual properties, and we provide comparisons to competing neural image reconstruction techniques. Our test results show that FoVolNet consistently achieves significant time saving over conventional rendering while preserving perceptual quality.}

\keywords{Volume data, volume visualization, deep learning, foveated rendering, neural reconstruction}

\CCScatlist{
  \CCScatTwelve{Computing methodologies}{Rendering};
  \CCScatTwelve{omputing methodologies}{Neural networks};
  \CCScatTwelve{Computing methodologies}{Volumetric models};
  \CCScatTwelve{Computing methodologies}{Image compression};
}

\teaser{
  \centering
  \includegraphics[width=\textwidth]{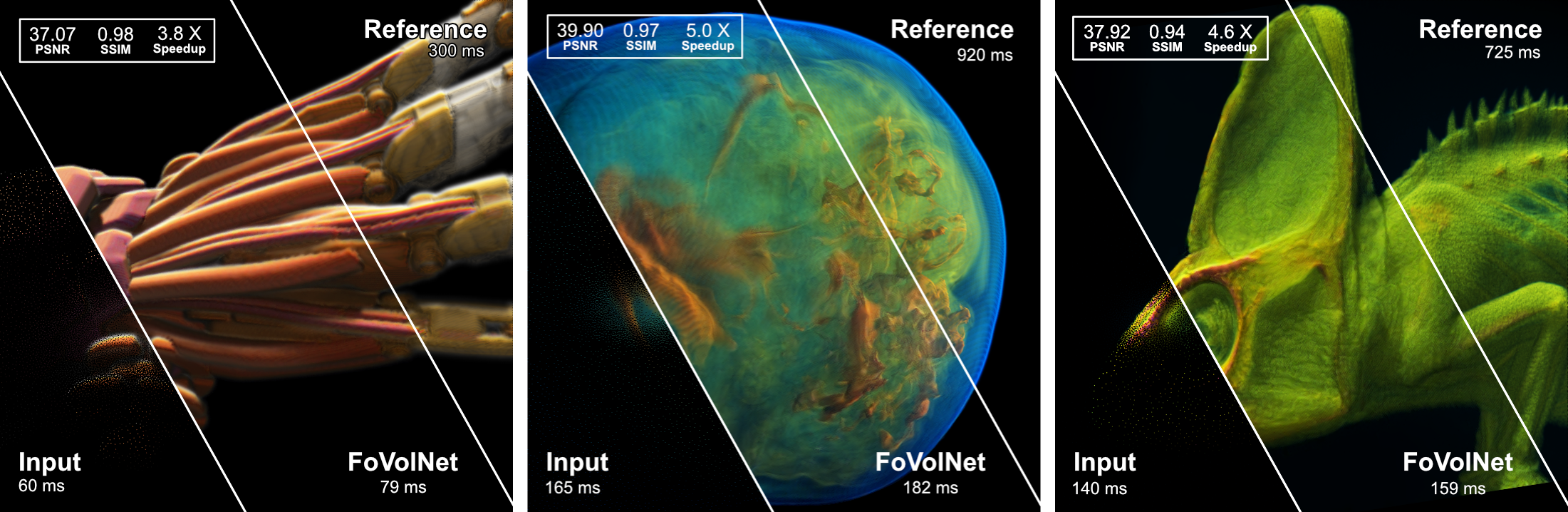}
  \caption{We propose a novel rendering pipeline for fast volume rendering using optimized foveated sparse rendering and deep neural reconstruction networks. FoVolNet can faithfully reconstruct visual information from sparse inputs. With FoVolNet, developers are able to significantly improve rendering time without sacrificing visual quality.}
  \label{fig:teaser}
}

\vgtcinsertpkg

\usepackage{amsmath,bm}
\usepackage{tabularx}
\usepackage{caption}
\usepackage{subcaption}
\usepackage{graphicx}
\usepackage{xcolor}
\usepackage{soul}
\definecolor{todotext}{RGB}{0,0,0}
\definecolor{todobg}{RGB}{255, 219, 168}

\begin{document}



\maketitle


\section{Introduction}
Since its beginnings, volume rendering has been an integral part of the scientific and biomedical visualization community. Over time, tremendous improvements have been made to the quality and performance of volume rendering algorithms. Yet, with advances in high-fidelity rendering comes increased computational cost. Many state-of-the-art techniques like path tracing or global illumination have outpaced the capabilities of consumer hardware, putting these techniques out of reach for interactive applications. There are also other relevant issues in this area, such as data management and storage. However, the visualization community has already produced various methods to mitigate these problems.


For instance, prior work~\cite{Shih2014ooc, Beyer2014largescale, Hardwiger2012petascale, Lundell2011OutofCoreMV} offers solutions for rendering extremely large volumes that do not fit the main memory of a system. Visualization of these data became viable through the introduction of streaming techniques such as out-of-core rendering which eliminate the need for the whole dataset to be present in memory.
Such methods are a means of emancipation. They help us depend less on specific characteristics of the data---in this case, its size.
When it comes to visual quality, state-of-the-art rendering techniques lack similar means. The higher the visual quality a technique produces, the more computational resources are generally needed to compute it. Although ongoing research has produced more efficient methods over the years, we are still bound by factors like the number of rays, sampling rates, or the type of illumination. Therefore, visualizing volume data using high-quality shading techniques at interactive framerates remains challenging---especially for demanding applications like immersive visualization. 




This work pries open the tight coupling between rendering technique and computational cost. We introduce FoVolNet---a complete volume rendering pipeline that aims to loosen the relation between technique and cost. By skipping the majority of screen-space pixel processing and replacing it with constant-time neural image reconstruction, we can achieve drastic performance improvements without sacrificing visual quality. We take inspiration from literature on foveated rendering and deep learning based image denoising. Prior work~\cite{Guenter2012, Stengel2016, Weier2016} has shown that taking the human visual system (HVS) into account when rendering data can yield excellent results for performance without perceptible quality loss. Utilizing the characteristics of the HVS is a crucial part of our design, as it allows us to concentrate computational resources. Accordingly, FoVolNet renders sparse images with dense foveated areas. A neural image reconstruction network restores the missing visual information, allowing us to skip the majority of screen-space pixel processing and replacing it with a constant-time inference step. This makes it possible to visualize volumes in high quality at a much lower computational cost than conventional rendering methods. In turn, this decoupling allows us to achieve faster and more consistent frame rates in various rendering setups. 

We conduct thorough tests involving the system's overall rendering performance, image quality, and other properties such as effective compression rate to evaluate our approach. The results show that FoVolNet faithfully reconstructs full frames at a fraction of the time it takes conventional methods to produce the same output.

\begin{figure*}[tbh]
  \centering
  \includegraphics[width=1\textwidth]{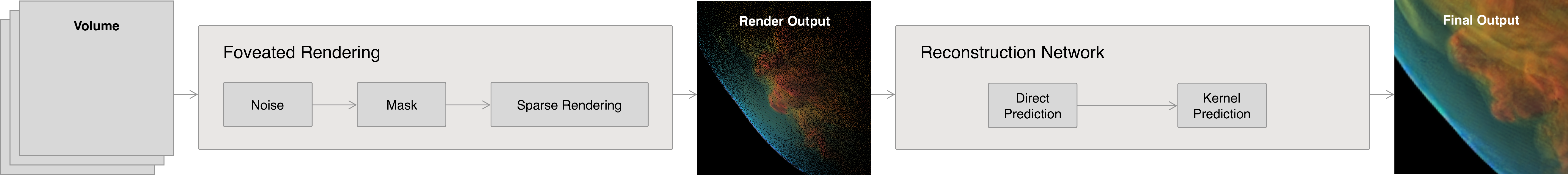}
    \caption{Overview of FoVolNet's components. The rendering pipeline loads a volume and renders it sparsely---saving time by skipping pixels in the periphery. Then, the sparse rendering is reconstructed by a neural network which takes constant time. The output of the rendering pipeline is the full-frame rendering.}
    \vspace{-0.15in}
  \label{fig:pipeline}
  
\end{figure*}

\section{Related Work}
Our work is related to topics in volume rendering, optical flow estimation, and deep learning methods in computer graphics and image processing. In this section, we discuss related works in these fields.

\subsection{Advanced Volume Rendering}

Optical models for computing advanced illumination effects (e.g., ambient occlusion, global shadows, multi-scattering) in volume rendering were first outlined by Max in the late 1990s~\cite{max1995optical, max2005local}. However, since these models are generally expensive to compute, a large body of work has focused on how performance can be improved.
The ambient occlusion model~\cite{diaz2010real, ropinski2008interactive, hernell2009local, ruiz2008obscurance, schott2009directional, vsolteszova2010multidirectional, kroes2015smooth} simulates the occlusion effect within a small neighborhood of the sample point, estimating the local extinction within a small spherical region. More recently, deep neural networks have been used to generate ambient occlusion effects for volume rendering~\cite{engel2020deep}. However, this model only accounts for local shadows and lacks cues for large-scale occlusions.
Computing global shadows require considering attenuation between light sources and the sample points. To efficiently compute global shadows, many different approaches have been proposed, including half-angle slicing~\cite{kniss2002interactive, kniss2003model}, plane sweep~\cite{sunden2011image}, shadow volume~\cite{ropinski2010interactive}, light volume~\cite{zhang2013fast}, or voxel cone tracing~\cite{shih2016parallel}. However, these methods still cannot calculate realistic multi-scattering effects.
More recently, the use of ray tracing presents a new trend of volume visualization algorithms that implement a highly realistic multi-scattering model ~\cite{kroes2012exposure, paladini2015optimization, paladini2015optimization, liu2016progressive, dappa2016cinematic}. 
By combining them with production ray-tracing software~\cite{wald2016ospray} and realistic BRDF classification techniques~\cite{igouchkine2017multi}, unbiased volume rendering can finally be achieved. However, these algorithms are computationally costly when high-resolution data---which requires a high sampling rate---and complex lighting conditions are combined. Thus, the rendering performance of these methods can quickly deteriorate. In this work, we lift a sizeable portion of the computational burden that such techniques incur on modern hardware. We reduce the screen space sample count and replace the skipped computations with a constant-time neural network inference step.

\subsection{Foveated Rendering}
Recent advances in eye-tracking technology and the market push towards augmented and virtual reality applications have intensified research in foveated rendering techniques. Computational power is crucial for high-fidelity rendering applications and most of today's immersive content. It is therefore of paramount importance to distribute resources efficiently. The capacity of the human visual system to perceive high levels of detail is limited to a relatively small focal area~\cite{Adler2011}. The fovea, which is the area of the visual field with the highest acuity, only makes up about $5.2$ degrees around the optical axis of the eye~\cite{Weier2017}. Foveated rendering approaches use that fact by focusing computational resources on these areas. Guenter et al.~\cite{Guenter2012} were one of the first to develop such an approach by rendering scenes in multiple levels of detail in concentric circles around the focus point. Later approaches~\cite{Stengel2016} use variable sampling rates that prioritize the focal area. Weier et al.~\cite{Weier2016} combine this approach with frame reprojection to reduce peripheral flickering.

\subsection{Deep Learning for Image Denoising}
One of the prominent uses of deep learning in the computer graphics field is image denoising. It is the process of refining noisy images, which are usually the result of Monte Carlo (MC) renderings with a low number of samples per pixel (SPP). A primitive approach to improving image quality is to increase SPP. However, this method requires significantly longer processing times per frame.
Recent work has leveraged deep learning to refine low SPP images without this computational overhead. Early approaches~\cite{Chaitanya2017, Bako2017} already achieve impressive results using CNNs. These works have established the two fundamental philosophies of image denoising in today's literature. On the one hand, there is direct prediction~\cite{Chaitanya2017}. A method to produce denoised images as a direct result of network inference. On the other hand, kernel prediction~\cite{Bako2017} approaches use CNNs to produce image filters. The denoising operation is performed by applying these filters to the input image in a separate step.

Subsequent work followed in these footsteps, furthering the potential of these two concepts. Wong et al.~\cite{Wong2018} introduce residual connections for direct prediction networks to improve single-frame image quality. To the same end, Xu et al.~\cite{Xu2019} and Lu et al.~\cite{Lu2020} conducted experiments on using adversarial networks~\cite{Goodfellow2014} to train direct prediction models. More recently, Hofmann et al.~\cite{Hofmann2020} have applied direct prediction to the domain of volume path tracing. Similarly, Weiss et al.~\cite{Weiss2022} explore the utility of direct prediction for reconstructing adaptive volume ray marching. Along with works like Kettunen et al.'s gradient-space denoising~\cite{Kettunen2019} and Wong et al.'s ResNet approach~\cite{Wong2018}, they investigate the effect of various auxiliary input features on final image quality. Following the kernel prediction path~\cite{Bako2017}, we see work by Vogels et al.~\cite{Vogels2018} who extend the approach by incorporating neighboring frames into training to facilitate temporal stability. Hasselgren et al.~\cite{Hasselgren2020} build on this notion, creating temporally stable image sequences using predictive adaptive sampling and temporal blending. Gharbi et al.~\cite{Gharbi2019} solve problems involving motion blur and depth-of-field using a kernel prediction approach with splatting.

Neural architectures used in these projects vary. However, there are certain identifiable trends. The U-Net architecture \cite{Ronneberger2015}, initially developed for medical image segmentation tasks, has proved to be a practical choice for denoising tasks. Many works \cite{Chaitanya2017, Hasselgren2020, Gharbi2019, Kettunen2019, Hofmann2020} base their design on the U-Net's image encoder-decoder principle. Extensions often include skip connections and recurrent feedback, which tend to increase image quality and temporal stability. Other works~\cite{Bako2017, Vogels2018, Wong2018, Xu2019, Lu2020} use more conventional CNN or RNN models. Interestingly, there is no apparent connection between chosen architecture (U-Net, CNN, RNN) and the denoising approach (direct prediction, kernel prediction). Several works~\cite{Xu2019, Lu2020, Hofmann2020} also use an additional critique network for their adversarial training. Named after its shape, the W-Net poses an extension to the U-Net and was introduced by Thomas et al.~\cite{Thomas2020}. It comprises two U-Nets in sequence. One is used for feature extraction, while the other serves to generate and apply convolutional filters to the input. This design facilitates optimization through selective quantization without significant image quality loss. For further reading on this topic, we refer the interested reader to Huo et al.'s survey on deep-learning-based image denoising techniques \cite{Huo2021}. 
Our network design is based on the W-Net architecture. We introduce a hybrid approach combining direct and kernel prediction to achieve the best results for sparse image inputs. This differs from conventional image denoising approaches in that missing visual information needs to be generated by the network. Therefore, pure kernel prediction networks are not suitable for this task as kernels only operate on existing pixel values. DeepFovea~\cite{KaplanyanDeepFovea} is the current state-of-the-art for such sparse frame reconstructions using solely direct prediction. Our approach translates this initial idea to the domain of scientific rendering and significantly improves visual quality and performance.


\subsection{Optical Flow}
Perceived motion in video sequences results from incremental changes in the positions of objects in a scene or by camera movement. Estimating the optical flow of elements in adjacent frames is an active area of research, and various approaches have been proposed in recent years. 

An early approach by Farnebäck et al.~\cite{Farneback2003} introduces a motion estimation algorithm that characterizes pixel neighborhoods as polynomials and uses those to find a mapping between frames. They propose a multi-scale approach that uses a priori motion estimation. This allows the algorithm to iterate and refine the estimation by considering differently sized search windows. This increases the robustness and quality of the results. Subsequent works~\cite{Sanchez2013, Plyer2016, LeBesnerais2005} tend to emulate this iterative, hierarchical approach. Most recently, Hanika et al.~\cite{Hanika2021Reprojection} have introduced a method based on this scheme. Unlike previous approaches, this algorithm sacrifices some quality in favor of speed. It also manages disocclusions gracefully.


For our work, we utilize Hanika et al.'s approach~\cite{Hanika2021Reprojection} to reproject frames during training. By warping a previous frame, we can gain more visual information about the current image, which can be used to construct a loss function that requires the network to match reprojected frames~\cite{KaplanyanDeepFovea}. This additional information helps increase image quality and supports retaining temporal stability between frames. 
\section{Methods}
FoVolNet is a complete raymarching volume rendering pipeline that is supported by a neural network (Figure \ref{fig:pipeline}). The overall rendering process consists of two critical steps. First, the volume needs to be rendered. Instead of rendering the whole frame, we selectively render a subset of pixels. A neural network is then used to reconstruct the full frame from this subset. The following sections contain details on our approach. There, we discuss the implementation and design of each stage of FoVolNet.

\subsection{Foveated Rendering}
We implement a ray marching system that facilitates sparse, foveated rendering. The renderer is implemented in CUDA and OptiX and supports global ray marched shadows. For the shadow computation we cast one shadow ray per sample step towards the light source using $\frac{1}{4}$ of the main sample rate. This renderer allows us to reduce rendering time as overall pixel density decreases. Our foveated rendering technique is based on binary sample maps generated from noise patterns that determine which pixels in screen space should be sampled by the volume renderer.


\begin{figure}[!htb]
    \centering

    \begin{subfigure}[t]{0.30\columnwidth}
        \centering
        \caption{Uniform Noise}
        \includegraphics[width=\linewidth]{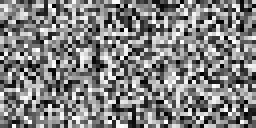} 
    \end{subfigure}
    \begin{subfigure}[t]{0.30\columnwidth}
        \centering
        \caption{Blue Noise}
        \includegraphics[width=\linewidth]{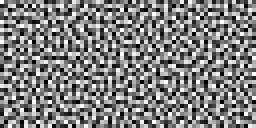} 
    \end{subfigure}
    \begin{subfigure}[t]{0.30\columnwidth}
        \centering
        \caption{STBN}
        \includegraphics[width=\linewidth]{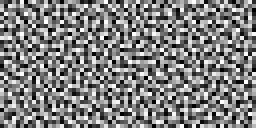} 
    \end{subfigure}

    \vspace{1mm}
    \begin{subfigure}[t]{0.30\columnwidth}
        \centering
        \includegraphics[width=\linewidth]{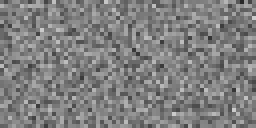} 
        \caption{Mean of Uniform}
    \end{subfigure}
    \begin{subfigure}[t]{0.30\columnwidth}
        \centering
        \includegraphics[width=\linewidth]{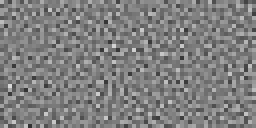} 
        \caption{Mean of Blue Noise}
    \end{subfigure}
    \begin{subfigure}[t]{0.30\columnwidth}
        \centering
        \includegraphics[width=\linewidth]{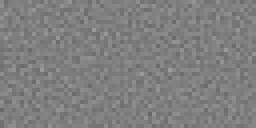} 
        \caption{Mean of STBN}
    \end{subfigure}

    \caption{We compare multiple noise patterns to create our sampling masks. The top row (a)-(c) shows the different types of noise. In the bottom row we show an 8-image sequence of patterns averaged across time to emphasize their temporal stability (d)-(f).}
  \label{fig:noise_patterns}
\end{figure}

\subsubsection{Noise Patterns}
The noise patterns used in this work (Figures \ref{fig:noise_patterns}, \ref{fig:sampling_masks} (left)) can be tiled seamlessly which allows us to cover an arbitrarily large frame. In our experiments, we tested noise map tile sizes that ranged from $16\times16$ to $256\times256$ pixels; however, there was no noticeable difference in the final image quality.

A comparison of different sources of noise is shown in Figure \ref{fig:noise_patterns}. Using noise sampled from a uniform distribution, like the one shown on the left, can result in energy spikes across the pattern. This is generally not desirable as it causes samples in the sampling map to be unevenly distributed. The temporal mean of uniform noise exhibits similar problems. Blue noise (Figure \ref{fig:noise_patterns} (b)) is rich in high frequencies and generally does not suffer from low-frequency energy spikes in the spatial domain. Its distribution closely models that of the visual receptors on our retina and is therefore ideal for creating perceptually unobtrusive sampling patterns. However, conventional blue noise suffers from temporal instability, as can be seen in Figure \ref{fig:noise_patterns} (e).

We use spatio-temporal blue noise (STBN)~\cite{wolfe2021stbn} to generate temporally stable sample patterns while preserving the perceptual advantages of conventional blue noise (Figure \ref{fig:noise_patterns} (c), (f)). As opposed to sequences of independent 2D blue noise or 3D blue noise patterns, STBN is not only blue in the spatial domain - every pixel is blue over time. This property is desirable since it helps our reconstruction network to produce stable and perceptually clean image sequences.

\subsubsection{Sample Maps}
\label{sec:samplemaps}
\begin{figure}[!htb]
  \centering
  \includegraphics[width=0.85\columnwidth]{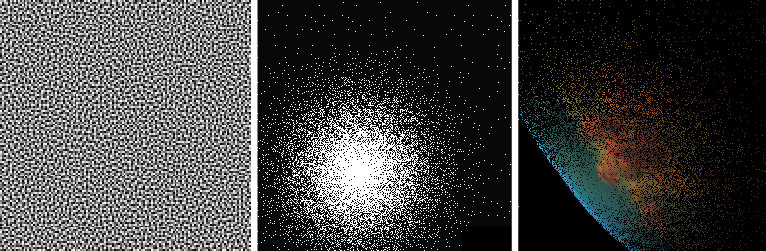}
    \caption{Sampling maps are generated using an STBN~\cite{wolfe2021stbn} (left). The area around the focal point is sampled more densely using an exponential fall-off to preserve details (middle). The volume is sparsely sampled using the sampling map (right).}
  \label{fig:sampling_masks}
\end{figure}

To generate a sample mask $M$, we compare the noise value $N(u,v)$ at position $u,v$ against a threshold $\tau$. If $N(u,v) < \tau$ we set $M(u,v)$ to $1$; otherwise, it is set to $0$. The value for $\tau$ can be adjusted to vary the sampling density. We define the density of the base noise pattern as $P_b(u,v)$. 
The foveated area is generated by modulating the value of $\tau$ using an exponential function around the focal point (Figure \ref{fig:sampling_masks} (middle)). Changing the variance $\sigma$, changes the size of the foveated area. The density of the foveated area is denoted by $P_f(u,v)$.

\begin{equation}
  P_f(u,v) = e^{-0.5 (f_x^2 + f_y^2) \sigma}
\end{equation}

\noindent where $f_x$ and $f_y$ are the current focal position. Combining both components, we calculate $\tau$ as follows.

\begin{equation}
    \tau (u,v) = (1 - P_b(u,v)) \cdot P_f(u,v) + P_b(u,v)
\end{equation}

\noindent With the resulting sampling mask, the volume is selectively rendered at all positions $(u,v)$ where $M(u,v) = 1$ (Figure \ref{fig:sampling_masks} (right)).

This process is repeated for every new frame that is rendered. To guarantee uniform sampling and maximize the amount of visual information that can be accumulated over time, the underlying blue noise maps are changed every time. Due to the computational complexity of blue noise generation, we use a pre-calculated series of 64 noise tiles. We loop the series to render frame sequences of arbitrary length.

Sampling only a small subset of rays can drastically reduce the computation time per frame. We define the maximum possible compression rate $C_{max}$ of the technique as follows.

\begin{equation}
    C_{max} = \frac{1}{h\cdot w} \sum_{u=0,v=0}^{h,w}{\tau (u,v)}
\end{equation}

\noindent where $h$ and $w$ are the spatial dimensions of the framebuffer. $P_b$ and $P_f$ are the probability of casting a ray at $u,v$ as described above.

\subsubsection{Rendering}
In a naive implementation of the renderer in OptiX and CUDA, invalid pixels would simply be discarded on the kernel level. 
However, using one kernel thread per full-size framebuffer pixel will yield only negligible performance gains. This is because GPU kernel calls are grouped in warps. Results are available only after all threads in a warp conclude. 
We develop two methods to circumvent this issue.

\subsubsection{Direct Sampling}

\begin{figure}[!htb]
  \centering
  \includegraphics[width=0.70\columnwidth]{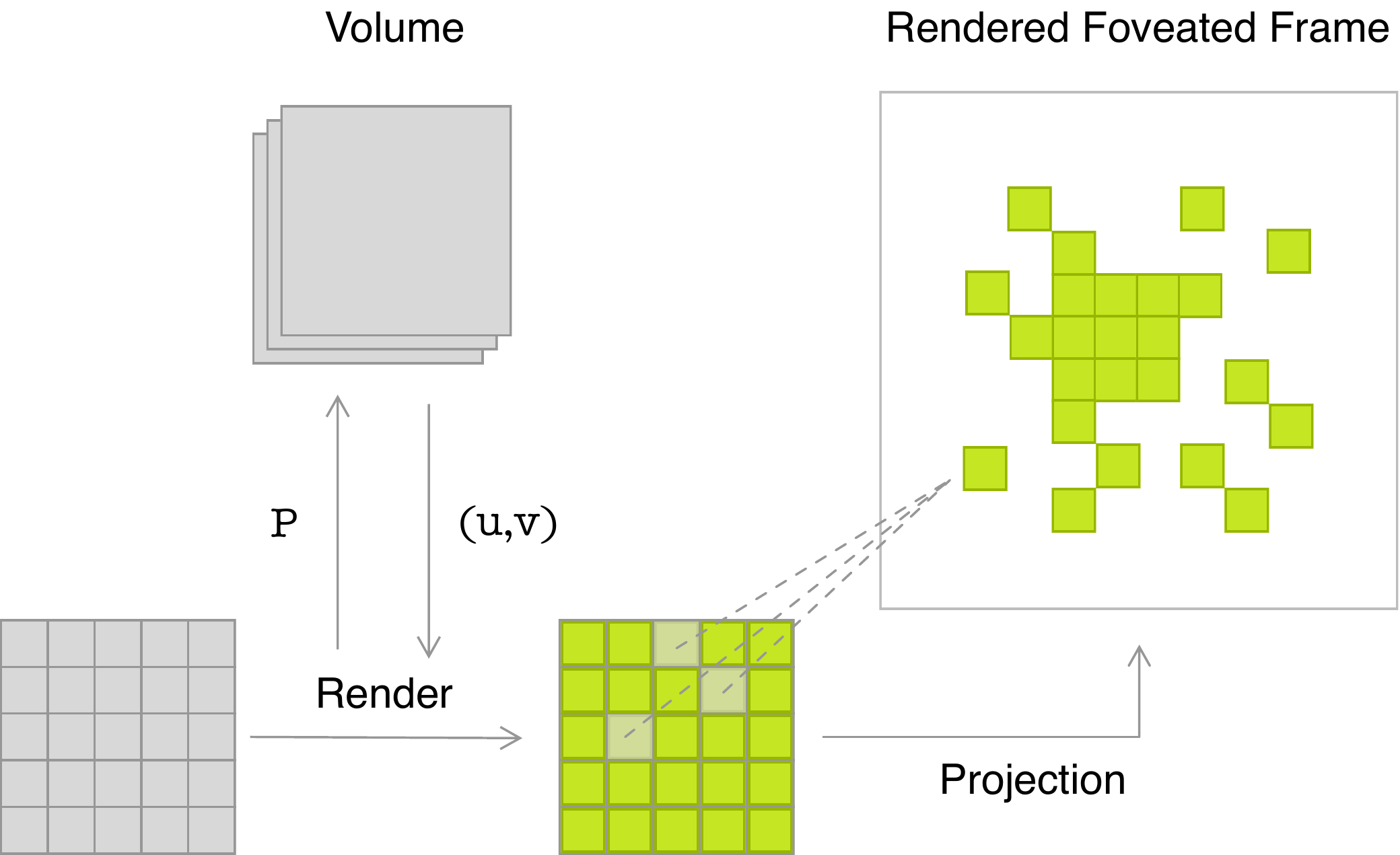}
    \caption{For direct sampling, we use a stochastic function $P$ to generate sample points. A small frame buffer is filled with samples that correspond to real locations on the full-size framebuffer. Color values are reprojected to the full frame.}
  \label{fig:direct_sampling}
\end{figure}

For this method, we create a stochastic function $P$ that incorporates both $P_b$ and $P_f$. It adapts over time as the noise pattern and the location of the fovea change. The function can be called to generate a position $u,v$ within the bounds of the framebuffer. Specific values for $u$ and $v$ are dependent on both probability functions. Therefore, it is more likely to generate a position in and around the foveated area. When rendering a new frame, we allocate a small framebuffer in which the number of pixels corresponds to the desired sampling compression $C_{max}$ which means that each kernel thread contributes calculations without potentially being discarded. For each entry in this buffer, the renderer queries $P$ to determine which pixel to render to the compact frame buffer. The values for $(u,v)$ relate to a pixel's position in the full-size framebuffer. Therefore, the result is a compact representation of the full frame. The contents of this compact buffer are projected back to their respective positions in the full-size framebuffer to generate the sparse image. Figure \ref{fig:direct_sampling} visualizes this process.

Direct sampling is relatively unintrusive in terms of pipeline integration. Instead of issuing a CUDA kernel run across the full image dimensions, we call it on a smaller buffer. Function $P$ acts as a proxy when accessing the screen-space coordinates in each GPU thread. The downside of this method is that $P$ does not reliably produce unique sampling positions. Although direct sampling performs better than the naive approach, we encounter duplicate coordinates quite frequently, making the renderer recompute existing samples, forfeiting potential compression. 

\begin{figure}[!htb]
  \centering
  \includegraphics[width=0.70\columnwidth]{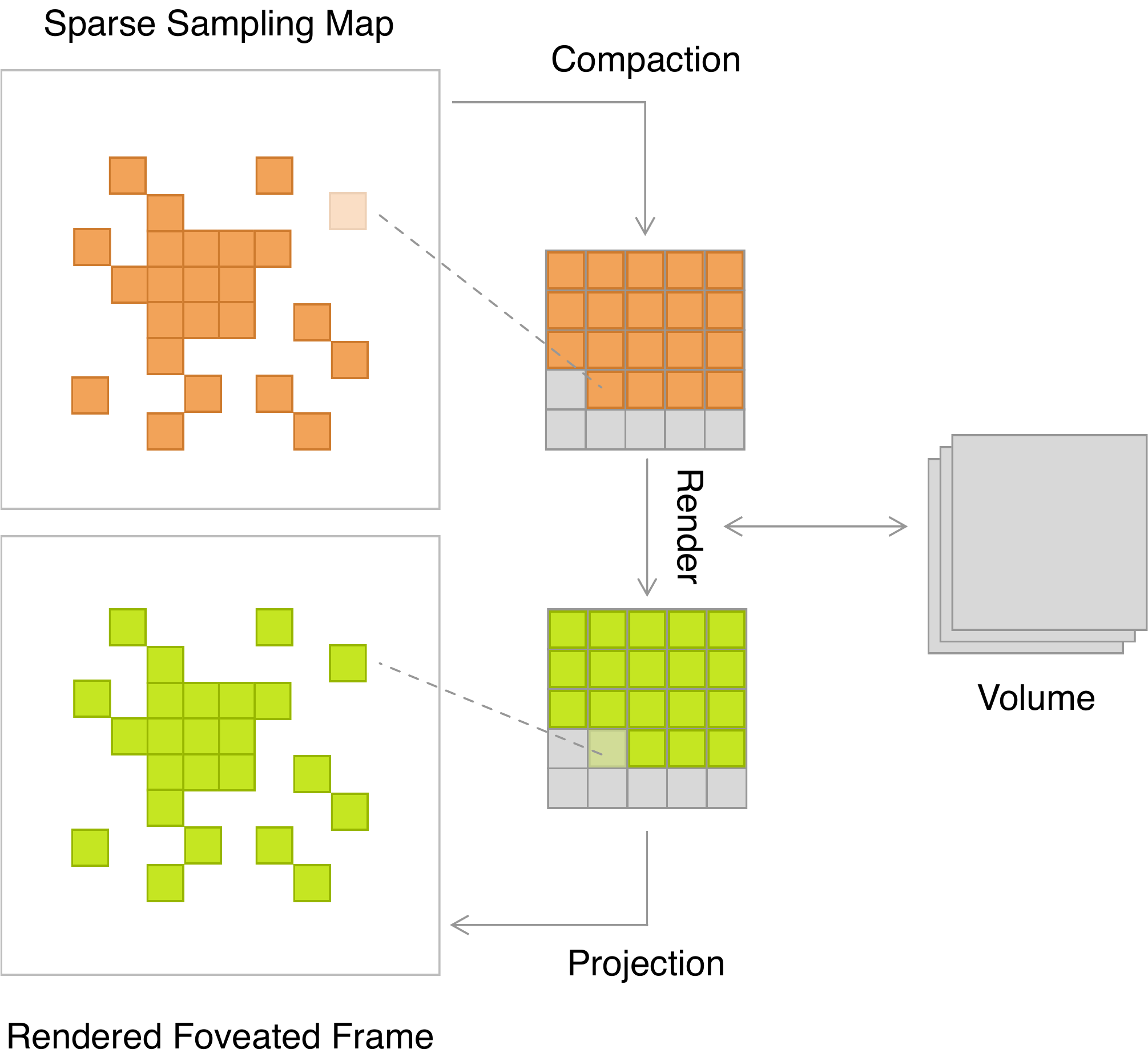}
    \caption{The renderer first creates a sampling map which is compacted into a small framebuffer. The volume is sampled according to each pixel's ray direction in the full-size frame. After rendering, the resulting color values are projected back to the initial frame.}
  \label{fig:stream_compaction}
\end{figure}

\subsubsection{Stream Compaction}
To alleviate the problems with direct sampling, we separate the sample map generation from the actual sampling. In the first step, the sampling map is generated using the approach described in Section~\ref{sec:samplemaps}. Next, a stream compaction algorithm is used to remove sparsity in the sampling map (Figure \ref{fig:stream_compaction}). This allows us to pack all valid pixels in the map into a smaller framebuffer similar to the direct sampling approach. The rendering is performed on this smaller framebuffer, and the same back-projection mechanism restores the full frame image (Figure \ref{fig:stream_compaction}).

The mask generation and compaction steps are implemented in CUDA to accelerate the process. Using this approach, we effectively circumvent the problems encountered in direct sampling and are therefore able to further approach the theoretical $C_{max}$.

\subsection{Reconstruction Network}
The core of FoVolNet is a deep neural network. We have developed a two-stage hybrid architecture that is based on the W-Net architecture~\cite{Thomas2020}. It draws ideas from both direct prediction and kernel prediction approaches. We specifically design this network to accommodate sparse frames with minimal input features. Images are reconstructed solely from RGB input---optical flow or other auxiliary features are not required. The network's components are described in detail here.

\begin{figure*}[!htb]
  \centering
  \includegraphics[width=0.95\textwidth]{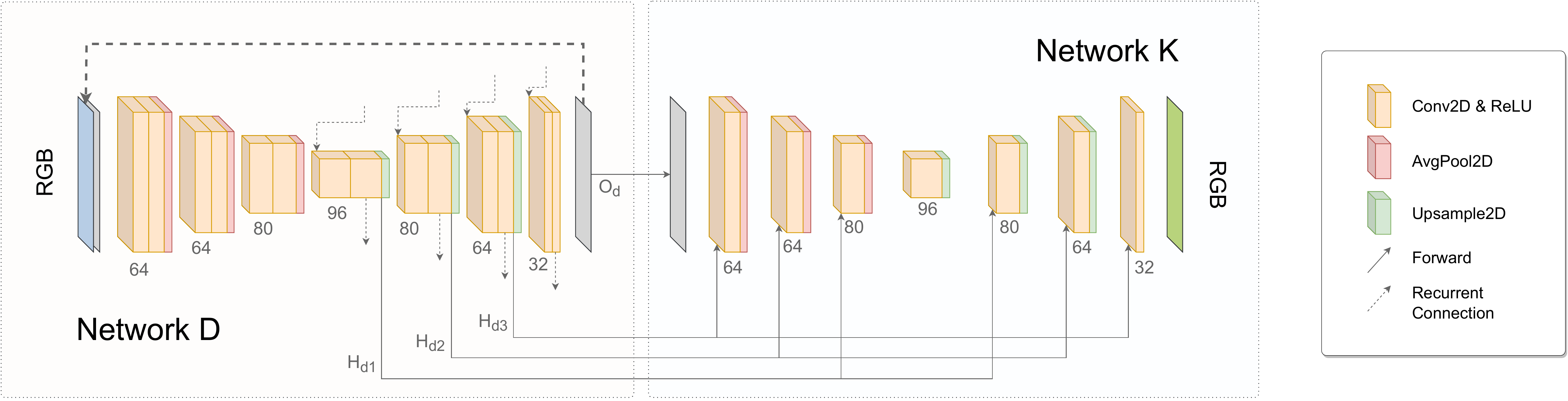}
    \caption{The reconstruction network comprises two U-Net stages $D$ and $K$. All \textit{Conv2D} layers are configured with a stride and padding of $1$ and no dilation. Network $D$ uses a $3\times3$ kernel size while $K$ uses $1\times1$ kernels. \textit{Upsample2D} layers use bilinear interpolation for filtering. Skip connection are not shown in favor of readability. Network $D$ performs coarse reconstruction through direct prediction while the second stage $K$ uses $D$'s hidden state to predict convolutional kernels which are subsequently applied to $D$'s output to produce the final frame.}
  \label{fig:network_architecture}
  \vspace{-0.15in}
\end{figure*}

\subsubsection{Overall Design}
The core idea of our design is to split the reconstruction process into two steps. The split is realized by running two networks in sequence (Figure~\ref{fig:network_architecture}).

This hybrid architecture employs both direct and kernel prediction. Without the initial reconstruction step, the kernel prediction method fails to perform due to the absence of rich pixel neighborhoods from which to draw visual information. On the other hand, performing only the direct prediction step would result in sub-par image quality, as we show in the evaluation.

\subsubsection{Direct Prediction: Coarse Image Reconstruction}
Network $D$ (Figure~\ref{fig:network_architecture}) reconstructs the image using direct prediction. That is, its output $O_d$ is directly interpretable as an image. This step reconstructs the overall features of the frame and fills the blank spots between valid pixels in the input. In addition to this, we preserve the decoder's hidden state $H_d$ for further processing.

\subsubsection{Kernel Prediction: Image Refinement}
In the second reconstruction step, network $K$ predicts convolutional kernels on multiple scales in both encoder and decoder stages. They are then applied in sequence to $O_d$. The kernel prediction stage takes the hidden states $H_d$ of $D's$ decoder stage (Figure~\ref{fig:network_architecture}) and forwards them to $K's$ convolution blocks in both the encoder and decoder stages. The convolution blocks are used to predict filter kernels from $H_d$. Network $K's$ input image is passed through the network by applying each block's predicted filter to the image in sequence. This process is analogous to the original W-Net filtering approach~\cite{Thomas2020}.

This step allows us to remove any remaining artifacts and blurriness that might result from direct prediction. Using the pre-filtered output $O_d$, we provide visual context for the filters to refine the image meaningfully. When using the original sparse image as input for this stage, we saw blotchy artifacts in areas with insufficient visual information to cleanly filter the image, and more densely sampled areas generally looked blurrier.





\subsubsection{Recurrence}
We introduce recurrent connections in multiple parts of the network to accumulate state. The aggregated information aids the reconstruction of temporally stable image sequences. 

Decoder blocks in network $D$ are connected by recurrent connections that pass down the block's output hidden state back to its input in the next training step. On a broader scale, the output $O_d$ of network $D$ is passed back to its input layer as part of the data of the subsequent run. Current and recurrent states are combined using a concatenation operation along the channel dimension, and appropriate up- or down-sampling is applied to make all inputs compatible for subsequent operations.

\subsection{Loss}
\begin{figure}[!htb]
  \centering
  \includegraphics[width=1.00\columnwidth]{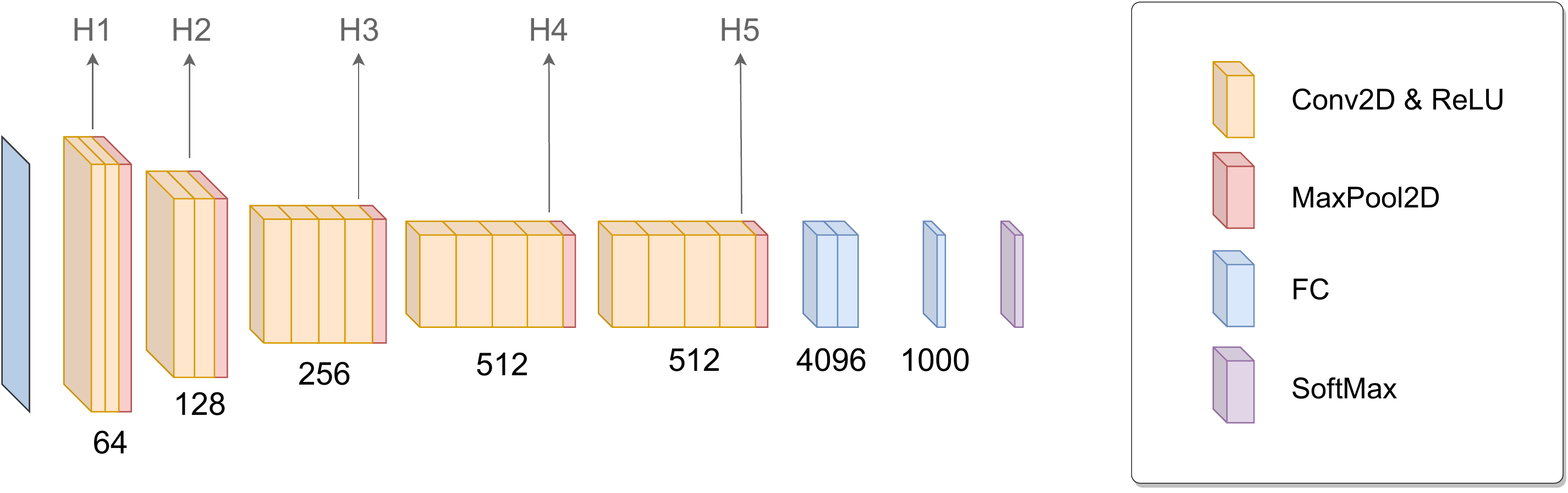}
    \caption{The hidden states H1-H5 of the VGG19 image classification network are used to measure the perceptual quality of our image reconstruction~\cite{zhang2018perceptual}.}
  \label{fig:lpips_network_architecture}
\end{figure}

Model optimization was performed using a linear combination of multiple loss components. We split the training loss into a spatial and a temporal component which we label $L_s$ and $L_t$, respectively. The losses are defined as follows.

\begin{equation}
    L = \lambda_s L_s + \lambda_t L_t
\end{equation}

\noindent where $\lambda_{s,t}$ denote the linear weights assigned to the loss. In our training we choose $\lambda_s = 0.8$ and $\lambda_t = 1.0$ which---given the losses' different magnitudes---weights spatial and temporal components at a ratio of 10:1. This balance of image quality and temporal stability was ideal for our trainings but might vary per training dataset. For $L_s$ we use a combination of $L_1$ and VGG19-based LPIPS perceptual loss~\cite{zhang2018perceptual} terms (Figure \ref{fig:lpips_network_architecture}). The temporal loss is a combination of $L_1$ loss and optical flow (OF) loss as used by Kaplanyan et al.~\cite{KaplanyanDeepFovea}.

\begin{equation}
    L_s = \lambda_1 LPIPS + \lambda_2 L_1 \quad\textrm{and}\quad L_t = \lambda_3 L_1 + \lambda_4 OF
\end{equation}


\noindent We choose $\lambda_1 = 0.9$, $\lambda_2 = 0.1$, $\lambda_3 = 1.0$, and $\lambda_4 = 0.1$ for the linear weights to equalize the components' magnitudes. Overall, the perceptual loss alone provides good reconstruction quality; however, adding a small $L_1$ term helps preserve some more high-frequency details.

\subsubsection{Spatial Loss}
During training, both $L_s$ and $L_t$ are applied across the whole series of images in a sequence. Both losses are computed for each time step and the model weights are updated once after a full sequence of loss values has accumulated. Given any pair of predictions $y_{p}$ and ground truth targets $y_{g}$ with $t$ total time steps, we calculate $L_s$ as follows.

\begin{equation}
    L_s(y_{g},y_{p}) = \sum_{i=0}^{t}{(1-e^{-0.5i}) \cdot (\lambda_1 LPIPS(y_{g_i}, y_{p_i}) + \lambda_2||y_{p_i} - y_{g_i}||_1)}
\end{equation}
Early images in the sequence are exponentially down-weighted to account for errors due to the lack of recurrent state at the beginning.

\subsubsection{Temporal Loss}
Using $L_s$ alone provides good single frame reconstruction quality. However, as previous studies~\cite{Hasselgren2020, Vogels2018, Chaitanya2017} have noted, it results in temporal flickering. Hasselgren et al.~\cite{Hasselgren2020} have shown that adding a simple $L_1$ term helps reduce temporal flickering drastically, but we have found their method to be prone to tearing artifacts when there is fast movement between adjacent frames. We add a small optical flow term, as used by Kaplanyan et al.~\cite{KaplanyanDeepFovea} to stabilize such cases. The first component forces the network to produce adjacent frames with finite differences similar to the output. The optical flow loss works by comparing the current frame against its predecessor. The previous frame is warped using the optical flow $\phi_{(i-1)\rightarrow i}$ with the warping operator $\omega$ to match the current frame. The network has to match this warped frame, leading to less tearing in the final output as consecutive frames become similar to their respective predecessors.
For a sequence of $t$ images $L_t$ is defined as follows.

\begin{equation}
\begin{split}
    L_t(y_{g},y_{p}) & = \sum_{i=1}^{t}{\sum_{j=0}^{i-1}{\lambda_3 ||(y_{p_i} - y_{p_j}) - (y_{g_i} - y_{g_j})||_1}} \\
    & + \sum_{i=1}^{t}{\lambda_4 ||y_{p_i} - \omega (y_{p_{i-1}}, \phi_{(i-1)\rightarrow i})||_1}
\end{split}
\end{equation}

In our training, the first loss term is defined over the whole sequence of prior frames. This way of constructing the loss emphasizes later image pairs in the sequence which entices training to use recurrent connections. On the other hand, the optical flow loss is only applied to a frame's direct predecessor as warping frames becomes harder and more prone to errors the farther they are apart temporally. We use Hanika et al.'s fast reprojection algorithm~\cite{Hanika2021Reprojection} to estimate optical flow between frames during training. For both components of $L_t$, we do not consider the first frame of the sequence as it has no viable predecessor.

\subsection{Model Precision \& Optimization}
Initially, we train the network using full 32-bit floating-point precision. However, the computational cost of running a full-precision network is often unnecessarily high. We truncate the network's weights to 16-bit half precision format as a first optimization step. This operation does not cause any noticeable performance loss.

We also experiment with post-training quantization on a pre-trained model. In this process, we initially train the model in full-precision mode. After this, the precision of the network is reduced, and training continues using half-precision. In contrast to similar works~\cite{Thomas2020, Jacob2018QuantizingEfficient}, we do not simulate integer quantization~\cite{Krishnamoorthi2018Quantizing}, as we observed drastic deterioration of image quality using this approach on our data. This is likely due to the sparsity of the input and the reliance on network $D$ to contribute to the final output instead of just extracting features.

Post-training quantization is realized using TensorRT~\cite{TensorRT}. Models are trained in PyTorch~\cite{PyTorch} and exported to ONNX format. We then transform the ONNX network to a CUDA inference engine using the tools provided by TensorRT~\cite{TensorRT}. In our trials, we experiment with different levels of optimization. Namely, we choose from different numerical precisions: float32, float16, int8, and mixed-mode. We compare different settings in Section~\ref{sec:evaluation}.

\subsection{Training}
We provide information about network configuration, hyperparameters, and details regarding the dataset used for training in this section. The training was performed on two NVIDIA Quadro RTX 8000 GPUs.

\subsubsection{Model Configuration}
Both sub-networks $D$ and $K$ of our reconstruction network are based on the U-Net design~\cite{Ronneberger2015}. Each network has four encoder and three decoder blocks. Skip connections connect the blocks. In network $D$, each block consists of two convolutional layers of equal depth, followed by a ReLU activation. On the other hand, each block in network $K$ only has a single convolution to predict the kernels. Both networks follow the same progression of block configurations which is defined as:

\begin{equation}
    e64-e64-e80-d96-d80-d64-d64
\end{equation}

\noindent where $e$ and $d$ denote encoder/decoder blocks followed by the convolution depth used for \textit{Conv2D} layers in the block. Blocks in the encoder stage conclude with an average pooling layer to downscale the image. Analogously, all except the final block in the decoder part up-sample their outputs. 

\subsubsection{Dataset}
FoVolNet is trained on short video sequences of several pre-rendered volume datasets. The data covers CT scans of humans, animals, mechanical parts, and large-scale simulation data from astronomy and material sciences (Table \ref{tab:datasets}). 

\begin{table}[!htb]
\centering
    \begin{tabularx}{\columnwidth}{l|X|l}
        \toprule
        \textbf{Dataset} & \textbf{Dimensions} & \textbf{Data Type} \\
        \midrule
        \midrule
        \textsc{Skull~\cite{skull}} & $256x256x256$ & uint8\\
        \textsc{Chameleon~\cite{chamaeleon,openscivis}} & $1024x1024x1080$ & uint8\\
        \textsc{Mechanical Hand} & $640x220x229$ & float32\\
        \textsc{Vortices 1~\cite{vortices}} & $128x128x128$ & float32\\
        \textsc{Vortices 2~\cite{vortices}} & $128x128x128$ & float32\\
        \textsc{Supernova~\cite{ornl}} & $432x432x432$ & float32\\
        \toprule
    \end{tabularx}
    
    \caption{Datasets used to train and evaluate FoVolNet. Using our ray marching renderer, each volume was rendered as a camera fly-through sequence around the object. The skull and chameleon datasets were used for evaluation and were not part of the training datasets.}
\label{tab:datasets}
\end{table}

We render the datasets at a resolution of $800x800$ using the previously described renderer. Video sequences consist of a continuous camera fly-through around the volume to cover most angles of the data. For training, the video dataset is sliced into 16-frame segments with no overlaps, and images are tiled at a resolution of $256x256$. We find that these spatiotemporal dimensions offer the best trade-off between training time and quality. Sequences with eight or fewer frames resulted in under-utilization of recurrent connections and, therefore, bad temporal coherence. Batches consist of $15$ such $16x256x256$ sequences. In total, the network is trained on $16000$ unique images. Our validation data consists of $1600$ unique images from the same datasets. Training, validation, and test datasets were split randomly at a 10:1:1 ratio.

\subsubsection{Data Augmentation}
We use data augmentation, which effectively increases the number of unique training sequences we provide to the network. During training, sequences of frames are subject to random augmentations to improve training effectiveness. Table~\ref{tab:augmentations} shows the list of augmentations used during training. Here, P(x) denotes the independent probability of each augmentation occurring for any given batch of data.

\begin{table}[!htb]
\centering
    \begin{tabularx}{\columnwidth}{l|X|l}
        \toprule
        \textbf{Name} & \textbf{Description} & \textbf{P(x)} \\
        \midrule
        \midrule
        Colors & Randomly permutes color channels & $0.6$\\
        Flip Horizontal & Flips whole sequence along y axis & $0.5$\\
        Flip Vertical & Flips whole sequence along x axis & $0.5$\\
        Grayscale & Converts RGB input to grayscale & $0.3$\\
        Static & Turns a real sequence into a number of static frames & $0.3$\\
        Padding & Pads the whole sequence by a random number of pixels & $0.1$\\
        \toprule
    \end{tabularx}
    
    \caption{List of image augmentations applied during training. All augmentations affect the whole sequence of images to not introduce any unwanted combinations of effects.}
\label{tab:augmentations}
\end{table}

\subsubsection{Hyperparameters}
During development, we experimented with different sets of hyperparameters to empirically determine the best settings for training. These include initial learning rate (LR), learning rate schedule, optimizer, weight decay, and length of training. For the final version, we use the following setup.

The LR is set to an initial value of $1.25e-3$, and a cosine annealing LR schedule is applied to gradually reduce the LR to a minimum value of $1e-8$. We use the Ranger~\cite{Ranger} optimizer with a weight decay set to $1e-2$ to stabilize training. The model usually converges at around epoch $60$-$80$. All trainings are stopped after $120$ epochs.

\section{Evaluation}
\label{sec:evaluation}
To show the potential of FoVolNet, we conduct several tests to evaluate different aspects of the system. All evaluations are conducted using our custom foveated rendering pipeline. 

\subsection{System Setup}
We use C++ backends for both PyTorch~\cite{PyTorch} and TensorRT~\cite{TensorRT} for inference. All evaluations were performed on an end-user machine with an Intel Core i7-6900K CPU with 128 gigabytes of RAM and an NVIDIA Titan RTX GPU. The system runs Ubuntu 20.04 LTS, and all parts of FoVolNet were developed and compiled on Linux. All frames are rendered at a resolution of $1280 \times 720$. We use our ray marching renderer and enable global shadows using a single light source per scene. Layer weights are truncated to fp16 precision, unless otherwise specified. All output was produced using images that were not in the training set.

Some results show comparisons to DeepFovea---the current state-of-the-art of foveated sparse frame reconstruction~\cite{KaplanyanDeepFovea}. We train a model of this architecture on our data in our training pipeline using hyperparameters as suggested by the authors.

\subsection{Inference Speed \& Pipeline Throughput}
To test the rendering throughput of our system, we use a fixed-path camera fly-through in our rendering pipeline. The camera path consists of a pattern of oscillating zoom with continuous rotation in two axes. This allows us to cover most aspects 
of the data using a small number of frames. Please refer to the supplemental video for more details.

\begin{figure}[!htb]
    \centering
    \includegraphics[width=1\columnwidth]{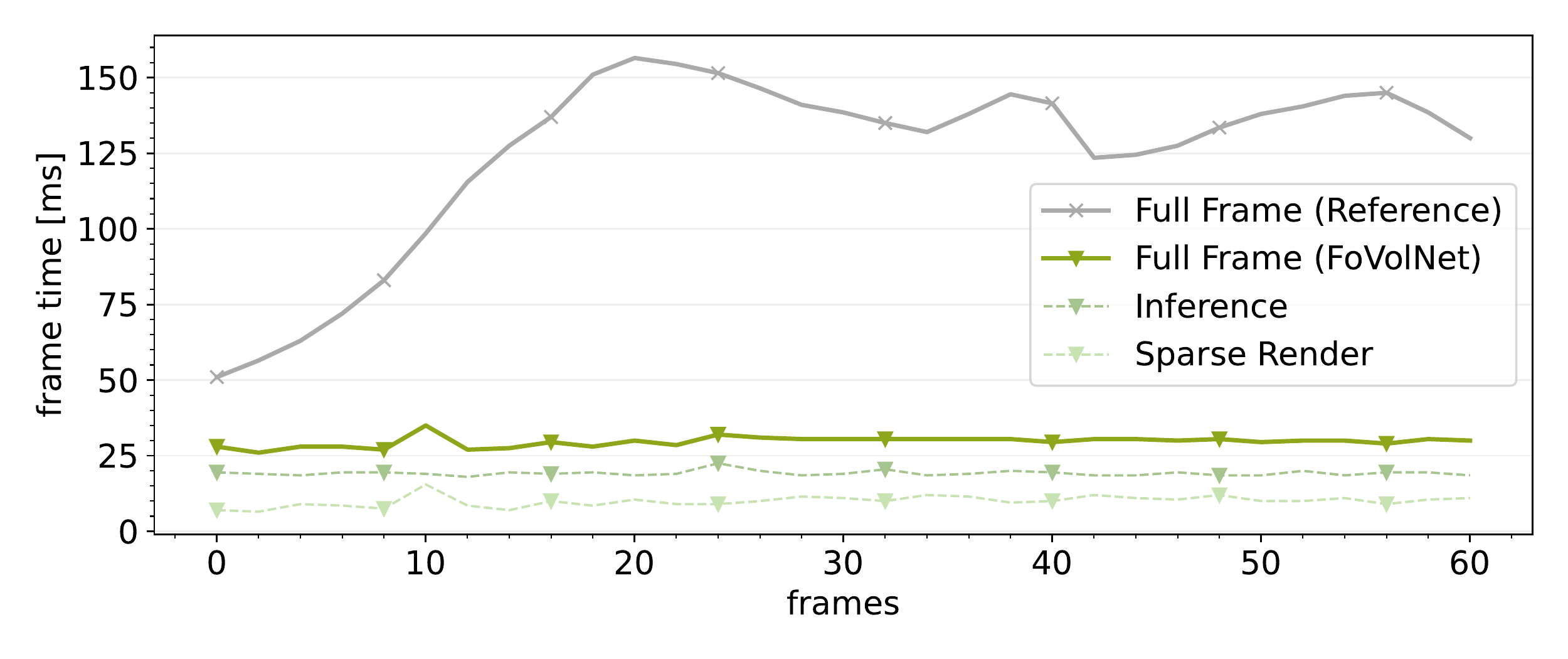}
    
    \hspace{7mm}
    \begin{subfigure}[t]{0.16\columnwidth}
        \centering
        \includegraphics[width=\linewidth]{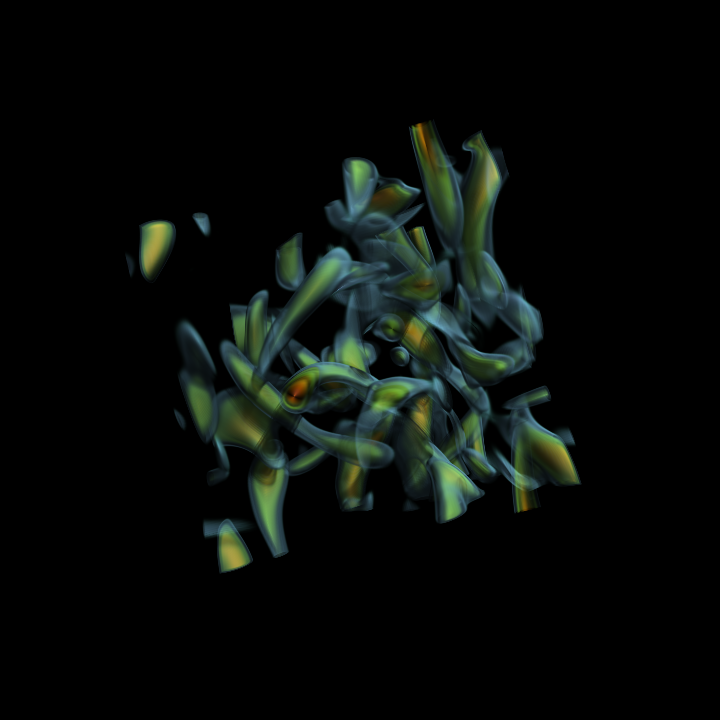} 
    \end{subfigure}
    \hspace{0.15\linewidth}
    \begin{subfigure}[t]{0.16\columnwidth}
        \centering
        \includegraphics[width=\linewidth]{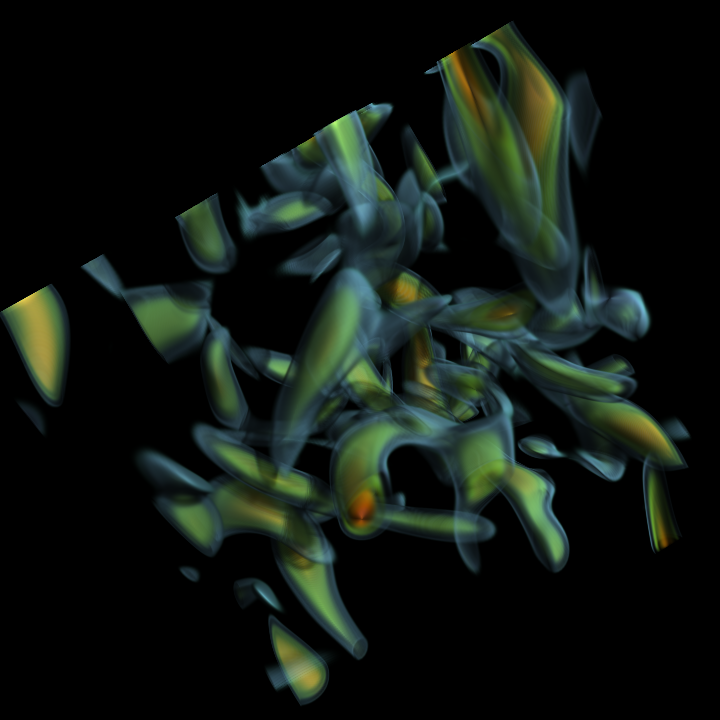} 
    \end{subfigure}
    \hspace{0.15\linewidth}
    \begin{subfigure}[t]{0.16\columnwidth}
        \centering
        \includegraphics[width=\linewidth]{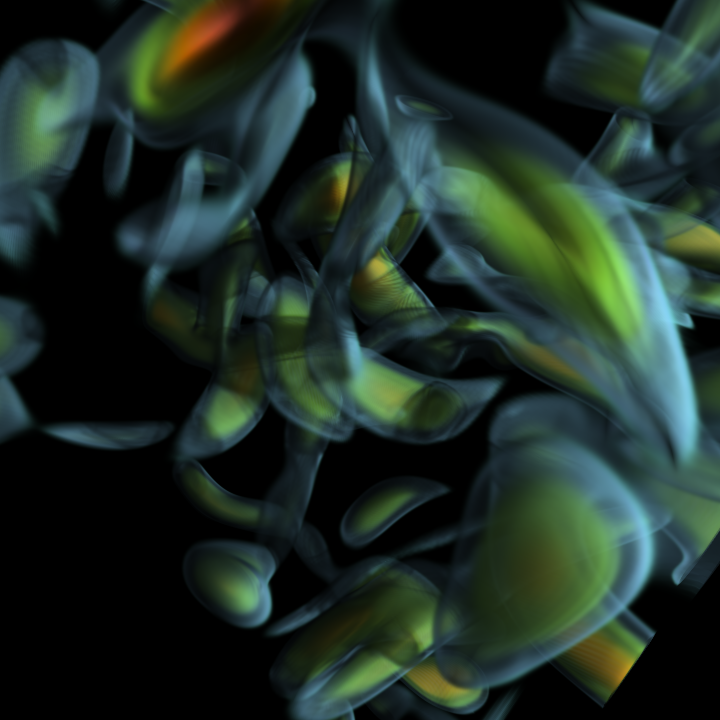} 
    \end{subfigure}
    
    \caption{The per-frame timings of different pipeline components during a camera fly-through of \textsc{Vortices 2}. We compare times for FoVolNet using fp16 precision with conventional DVR as the reference. Thumbnails show the camera position at that specific point in the run.}
      \label{fig:result_vortices_timings}

\end{figure}

A comparison of conventional raymarching with FoVolNet is shown in Figure \ref{fig:result_vortices_timings}. We perform the fly-through mentioned above on the Vortices 2 dataset. Baseline rendering time is drastically reduced due to sparse, foveated sampling of the volume. A constant, scene-independent inference time adds to the total frame time. The result is a sequence of fast and stable frame times that is less dependent on camera angle or scene configuration. As the thumbnails in Figure \ref{fig:result_vortices_timings} suggest, the more screen space is occupied by data, the larger the benefit of using our technique. However, as the left-most image shows, we achieve roughly two times faster end-to-end rendering performance even from a far-away viewing position.

\begin{figure}[!htb]
    \centering
    \includegraphics[width=1\columnwidth]{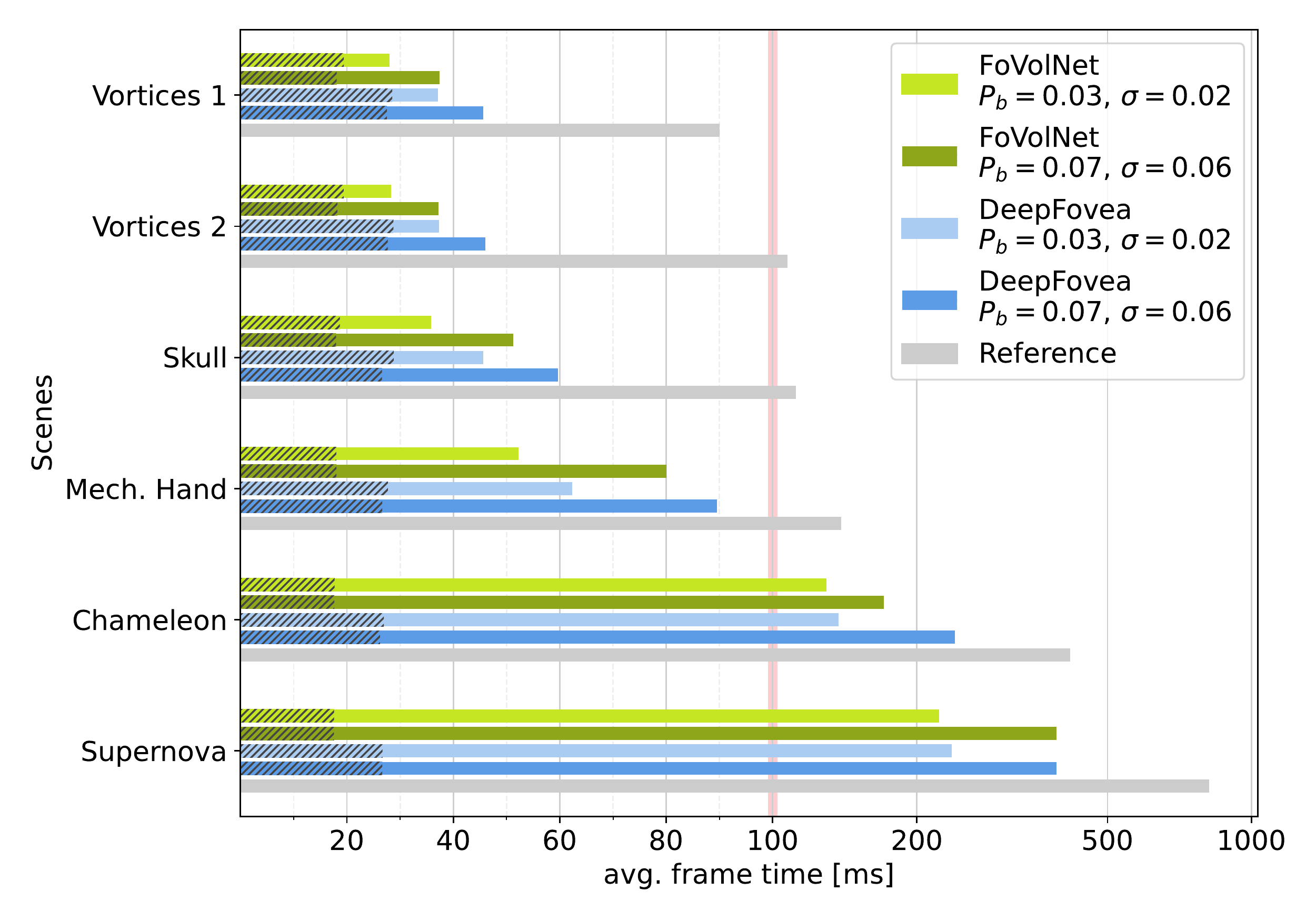}
    \vspace{-0.2in}
    \caption{Average frame times from fly-through renderings of different datasets. Each dataset was rendered for 500 frames. The camera movement was framerate-independent. We compare against DeepFovea~\cite{KaplanyanDeepFovea} as specified by the authors. Note that the x-axis scales logarithmically past frame 100 to accommodate long frame times.}
    \label{fig:result_avg_timings}
\end{figure}

A more comprehensive analysis of rendering performance is shown in Figure \ref{fig:result_avg_timings}. Here, we compare against DeepFovea~\cite{KaplanyanDeepFovea}. The sequences were created at two different quality settings, which differ in their configuration of sampling density. The hatched parts of the deep learning based runs indicate the inference times. We report the resulting average end-to-end speedups for all datasets in Table \ref{tab:result_avg_timings_speedup}. Note that the fly-through sequences are all composed of roughly equal parts far-away and close-up viewing positions. This is due to the oscillating zoom of the camera. Therefore, our results represent speedups that can be expected in the average case. However, if the data is viewed at a reasonably close angle like shown in Figure~\ref{fig:teaser}, speedups are generally much higher.

\begin{table}[h]
    \caption{Relative speed-up times when compared to the baseline raymarching renderer. The data is based on that shown in Figure \ref{fig:result_avg_timings}. $P_b$ is given and $P_f$ was calculated using the listed $\sigma$ values.}
    \centering
    \begin{tabularx}{\columnwidth}{l|X|X|X|X}
    \toprule
        \textbf{Dataset} & \textbf{FoVolNet $P_b=0.03$ $\sigma=0.02$} & \textbf{FoVolNet $P_b=0.07$ $\sigma=0.06$} & \textbf{DeepFovea $P_b=0.03$ $\sigma=0.02$} & \textbf{DeepFovea $P_b=0.07$ $\sigma=0.06$} \\
    \midrule
    \midrule
        \textsc{Vort. 1}      & $3.21\times$ & $2.40\times$ & $2.42\times$ & $1.98\times$ \\
        \textsc{Vort. 2}      & $3.80\times$ & $2.89\times$ & $2.88\times$ & $2.34\times$ \\
        \textsc{Skull}           & $3.12\times$ & $2.19\times$ & $2.45\times$ & $1.88\times$ \\
        \begin{tabular}[l]{@{}l@{}}\textsc{Mech.}\\\textsc{Hand}\end{tabular} & $2.66\times$ & $1.74\times$ & $2.23\times$ & $1.55\times$ \\
        \textsc{Cham.}       & $3.22\times$ & $2.45\times$ & $3.05\times$ & $1.73\times$ \\
        \textsc{Supern.}       & $3.67\times$ & $2.09\times$ & $3.44\times$ & $2.08\times$ \\
        \textbf{Overall}& $\bm{3.28\times}$ & $\bm{2.29\times}$ & $\bm{2.75\times}$ & $\bm{1.93\times}$\\
    \toprule
    \end{tabularx}
    \label{tab:result_avg_timings_speedup}
    \vspace{-0.1in}
\end{table}

\subsection{Image Quality}


\begin{figure*}[!htb]
  \centering
  \includegraphics[width=1\textwidth]{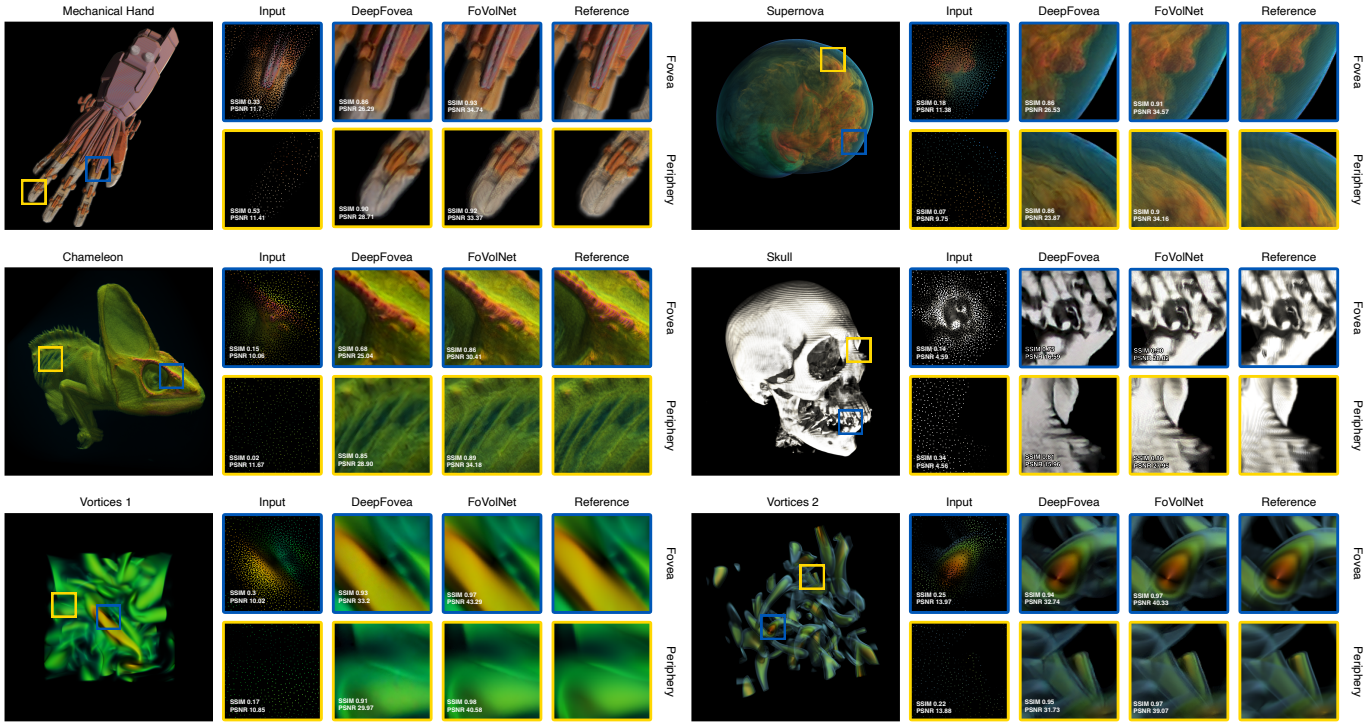}
    \caption{Visual comparison of reconstruction quality using our method. For each dataset, we show the area around the fovea in blue and a part of the periphery in yellow. All images were generated with $P_b = 0.03$ and $\sigma = 0.02$ for $P_f$.}
  \label{fig:quality_matrix}
  \vspace{-0.2in}
\end{figure*}

Final image quality is at least as important as inference speed when it comes to image reconstruction. For all datasets, we calculate structural similarity (SSIM) and peak signal to noise ratio (PSNR) on single frames and image sequences. The image matrix in Figure~\ref{fig:quality_matrix} shows results for both foveated areas and the periphery on single frames. Our architecture can reconstruct fine details even in peripheral areas of the frame. Notice how pure direct prediction methods like the recurrent U-Net used in DeepFovea~\cite{KaplanyanDeepFovea} fail to preserve high-frequency details as the number of samples decreases.

\begin{figure}[!htb]
    \centering
    \includegraphics[width=0.85\columnwidth]{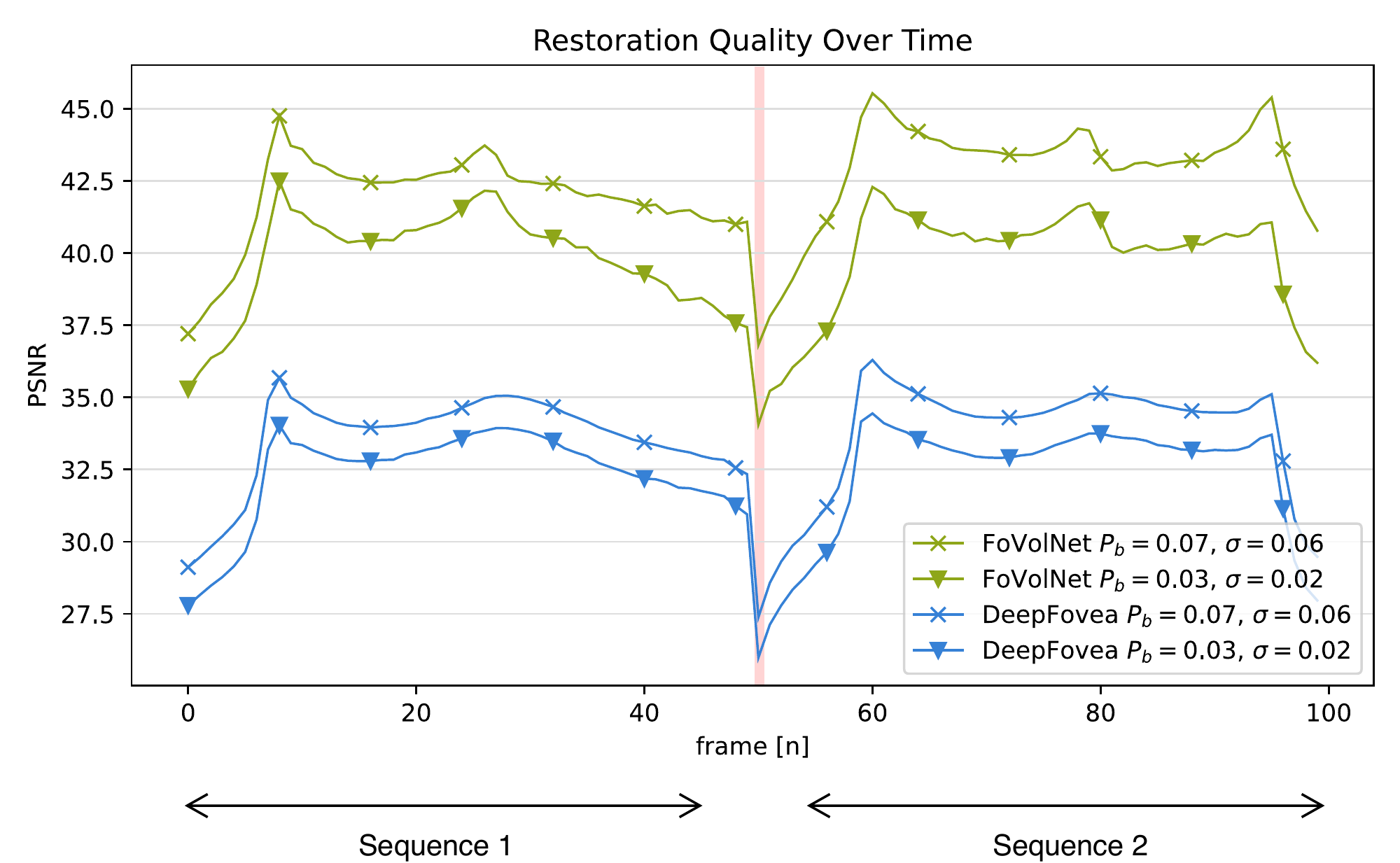}
    \caption{Still-frame PSNR over the course of two 50-frame clips from the \textsc{chameleon} dataset.}
    \label{fig:result_quality_over_time_psnr}
    \vspace{-0.2in}
\end{figure}

\begin{figure}[!htb]
    \centering
    \includegraphics[width=0.85\columnwidth]{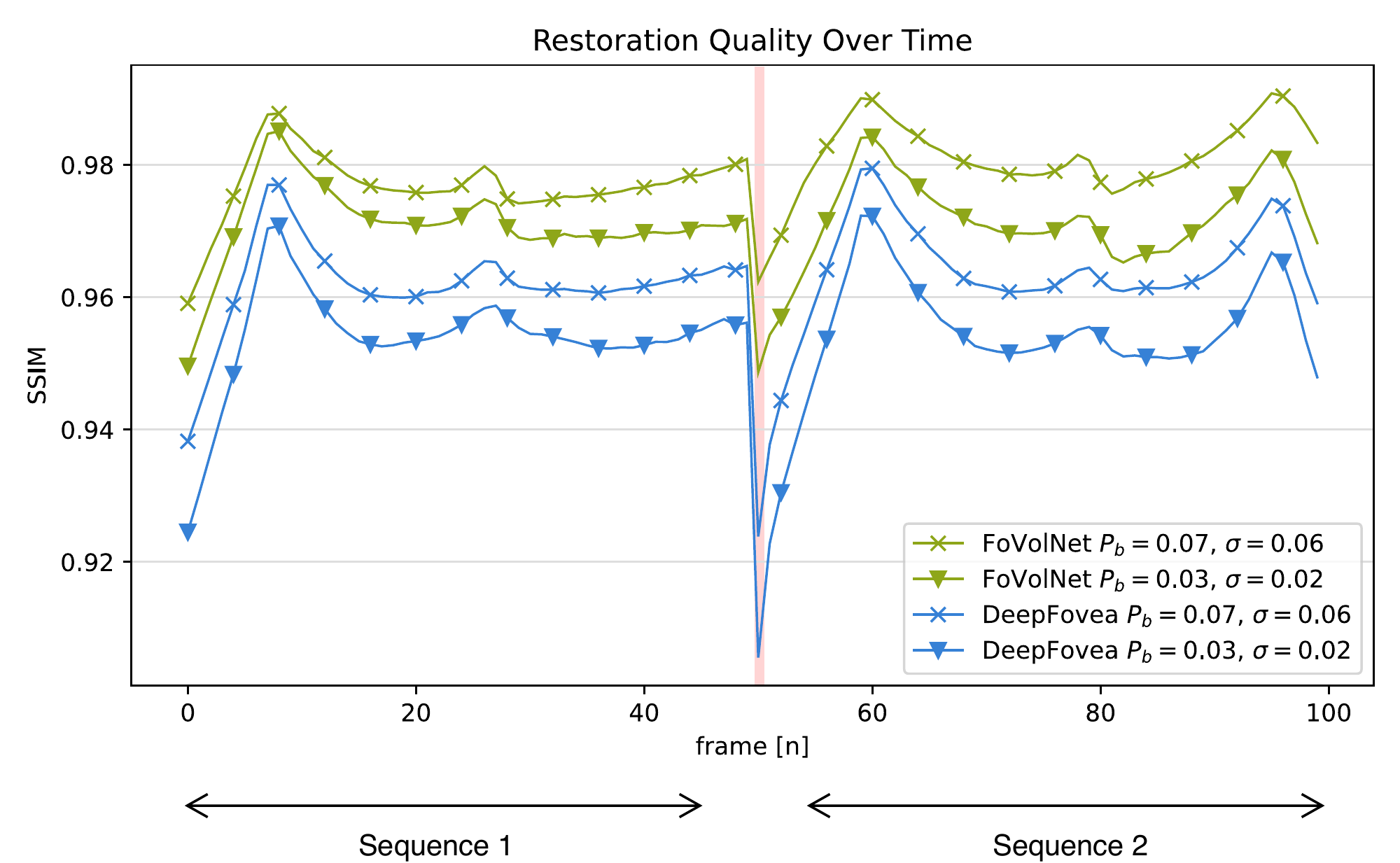}
    \caption{Structural similarity (SSIM) over the course of two 50-frame clips from the \textsc{chameleon} dataset.}
    \label{fig:result_quality_over_time_ssim}
\end{figure}

For the video analysis, we create a camera fly-through video of the \textsc{chameleon} dataset. We split the video into multiple 50-frame sequences (of which we show two) with a jump-cut between them. Figures~\ref{fig:result_quality_over_time_psnr} and~\ref{fig:result_quality_over_time_ssim} show the reconstruction quality over time. A red line indicates the cut. Both FoVolNet and direct prediction improve their quality over a short ramp-up period in which state is accumulated. Similarly, the network needs several frames after the jump cut to recover full quality. However, the hybrid architecture outperforms direct prediction by a constant offset. 

\subsection{Temporal Stability}
\begin{figure}[h]

    \centering
    \includegraphics[width=0.85\columnwidth]{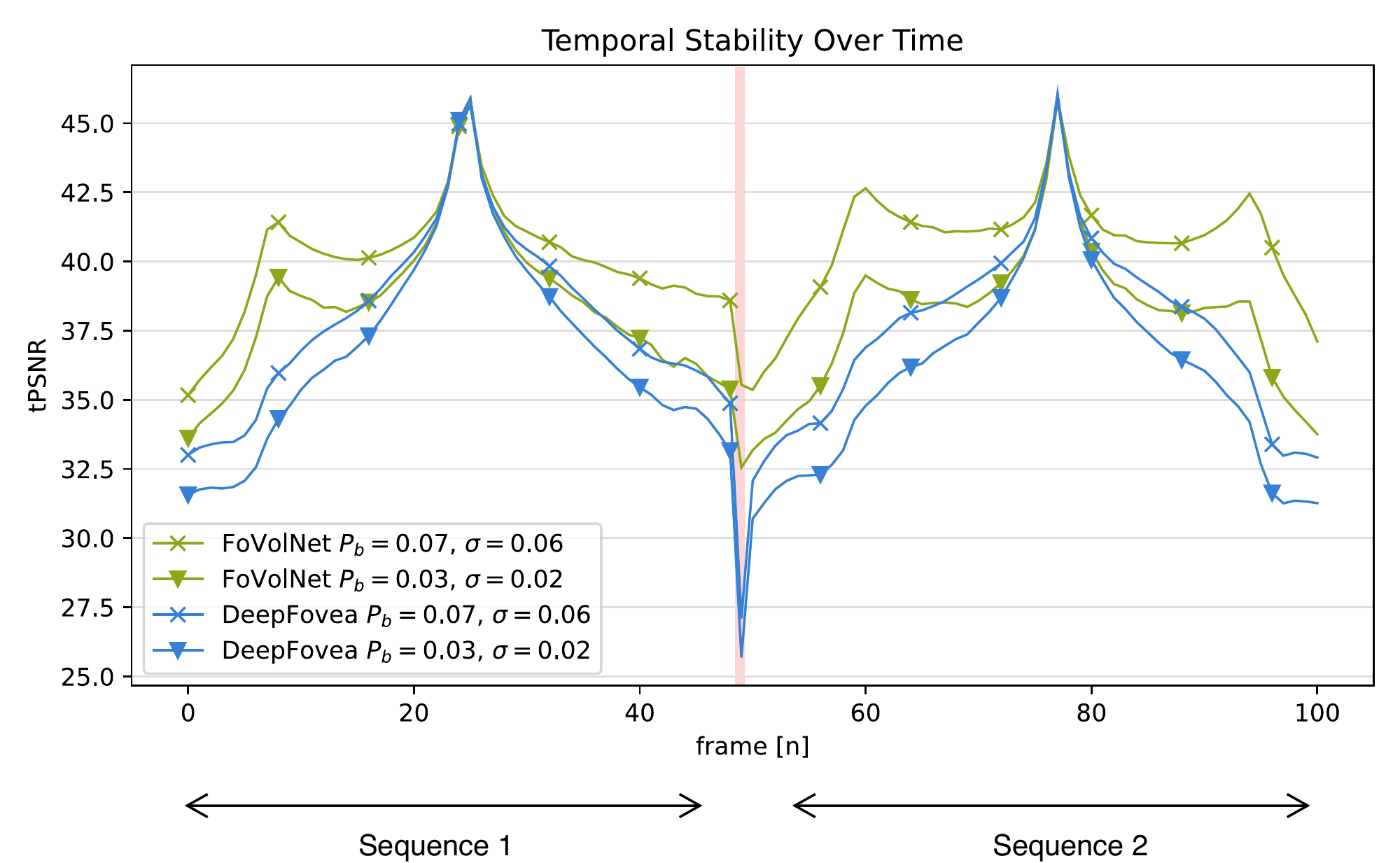}
    \caption{Temporal stability as measured by tPSNR~\cite{Hasselgren2020} over the course of two 50-frame clips from the \textsc{chameleon} dataset.}
    \label{fig:result_temporal_stability_over_time}
\end{figure}

The quality of temporal coherence is evaluated on the same fly-through clips. The sequences were constructed so that they each start and end in fast camera movement while slowing down towards the middle. This three-act setup allows us to see how the network uses accumulated state to retain temporal consistency when there is (1) plenty of movement with little prior state, (2) little movement but lots of state, and lastly, (3) lots of movement and lots of state. We compute the temporal PSNR (tPSNR) as used by Hasselgren et al.~\cite{Hasselgren2020}. This value is the PSNR of finite differences between frames. Instead of computing the PSNR on the image itself, we compute it on the difference between the current and previous frame. The aforementioned fast-slow-fast pattern is reflected by data in Figure~\ref{fig:result_temporal_stability_over_time}. It shows that FoVolNet is able to retain stability throughout most of the sequences. Both models achieve peak quality when there is little movement. This is unsurprising since, without movement, the network acts as a simple accumulation buffer. However, FoVolNet is able to retain quality under much faster movement than direct prediction---especially in Phase (3) when there is enough state available to the layers. Please refer to the supplemental material for the source clips.

In addition to this metric, we provide just-objectionable-difference (JOD)~\cite{jodscore} scores for the whole video. This score is similar to just-noticeable-difference (JND), but instead of quantifying the difference between pairs of images, it is better suited to compare multiple degraded images to the reference. That means that while the results of different reconstruction methods might look degraded in different ways, they will still have similar JOD scores as they are equally different from the ground truth. The data was produced using FovVideoVDP~\cite{fovvideovdp}. We use the default settings for a 4K screen viewed under office light levels. Detailed settings were chosen as follows: 75.4 [pix/deg], Lpeak=200, Lblack=0.5979 [cd/m\textsuperscript{2}], non-foveated, (standard\_4k). Data is produced for the video at different quality settings as shown in Table~\ref{tab:result_temporal_stability_jod}. Note that higher scores are better, with 10 being the maximum score. 

\begin{table}[h]
    \caption{Just-objectionable-difference scores~\cite{fovvideovdp} of reconstruction output from the \textsc{chameleon} dataset computed at different thresholds.}
    \centering
    \begin{tabularx}{\columnwidth}{l|X|X}
    \toprule
        $\boldsymbol{\tau}$ & \textbf{JOD Score FoVolNet} & \textbf{JOD Score DeepFovea} \\
    \midrule
    \midrule
        0.10 & 9.35 & 8.53\\
        0.07 & 9.28 & 8.47\\
        0.03 & 8.86 & 8.21\\
        0.01 & 8.26 & 7.51\\
    \toprule
    \end{tabularx}

    \label{tab:result_temporal_stability_jod}
        \vspace{-0.15in}
\end{table}

\subsection{Model Precision}
\begin{figure}[!htb]
    \centering
    \includegraphics[width=1\columnwidth]{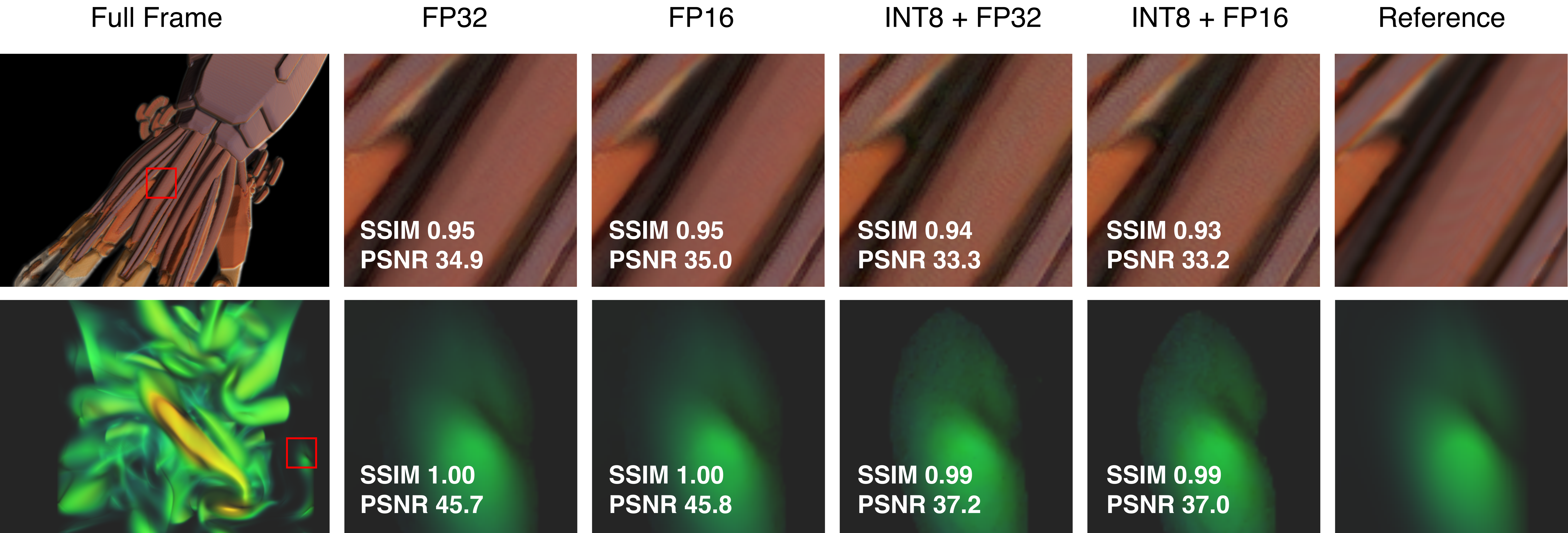}
    \caption{Comparison of reconstruction quality at different model precisions. Differences are most apparent on homogeneous surfaces and along subtle color gradients.}
    \label{fig:result_precision}
\end{figure}

During training, we configure the weights to use full 32-bit precision. By default, this level of precision is retained during inference. However, reducing the precision of certain weights in the network can drastically improve the performance during inference. We examine the effects that such adjustments have on image quality in practice (Figure~\ref{fig:result_precision}).

In our tests, quantization artifacts were most apparent on homogeneous surfaces, subtle gradients, and transparent regions. Two examples are shown in Figure~\ref{fig:result_precision}. The difference in quality is especially apparent in the lower dataset shown in Figure~\ref{fig:result_precision}. The fading color towards the top shows a much more abrupt cut-off in quantized precisions (int8 and mixed-precision int8/fp16) compared to their unquantized counterparts (fp32 and fp16). The brightness of the lower frame was increased to emphasize the subtle differences. There was no noticeable difference between the non-quantized precisions fp32 and fp16. 

\subsection{Effective Compression Rate}
\begin{figure}[!htb]
    \centering
    \includegraphics[width=0.95\columnwidth]{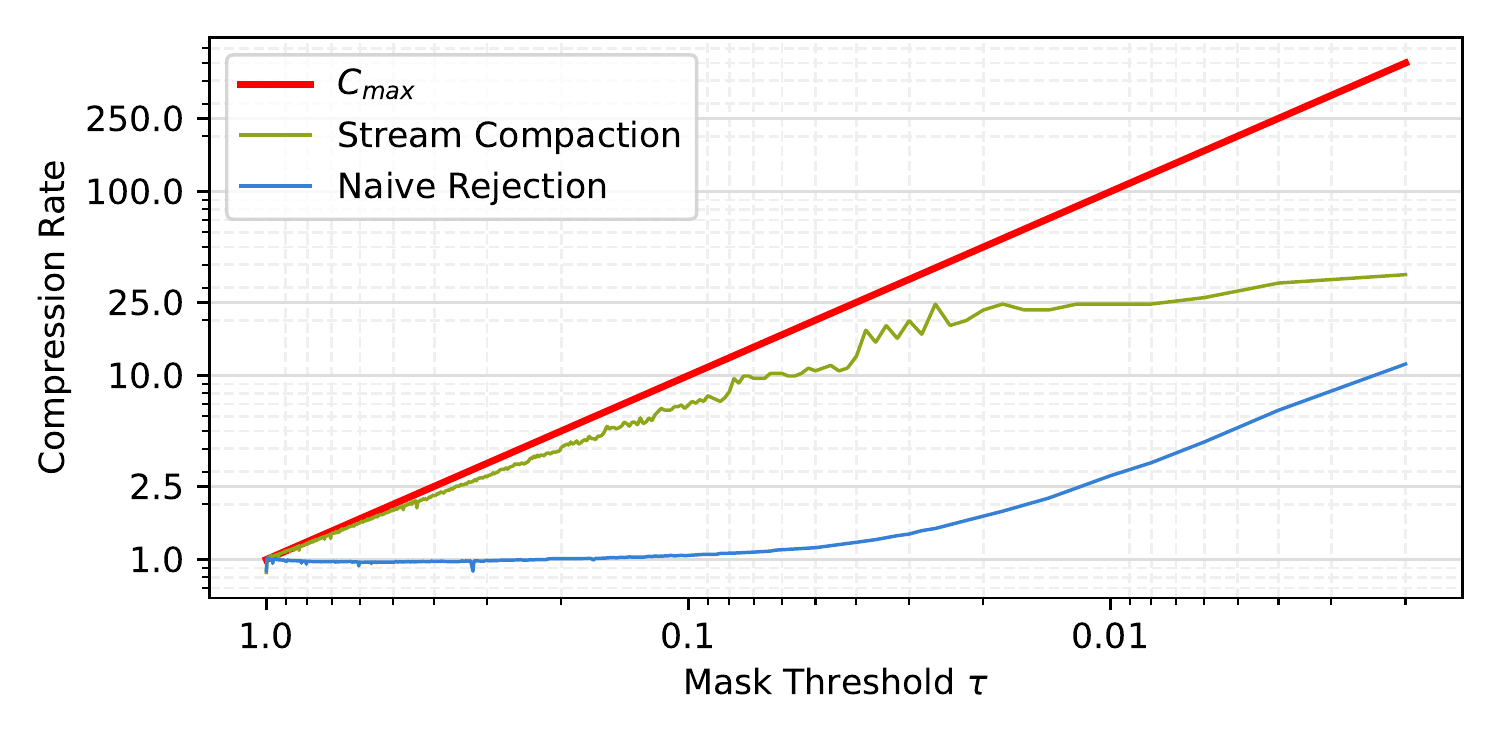}
    \vspace{-0.25in}
    \caption{Compression efficiency of our stream compaction rendering technique. Data was recorded for the whole value spectrum of sampling threshold $\tau$. A curve for $C_{max}$ shows the maximum achievable efficiency at any given threshold value.}
    \label{fig:result_compression}
    \vspace{-0.2in}
\end{figure}

Reducing the number of total pixels that the volume needs to be sampled at immediately affects rendering performance. In the ideal case, a sparse rendering algorithm would achieve rendering times that scale linearly with the number of pixels. We termed this ideal case $C_{max}$---the maximum possible rendering performance at any given sparsity level. To test how well our stream compaction sparse renderer performs, we record frame times along the whole spectrum of sparsity as represented by $\tau$ ranging from $1.0$ (full frame) to $0.0$ (no samples). Data is recorded for both the stream compaction method and a naive rejection approach which simply skips computations for certain threads on a full-frame kernel run. The results of this test are shown in Figure~\ref{fig:result_compression}.

The data shows that our compaction method maps well to $C_{max}$. As $\tau$ approaches sparsity rates of around $10\%$, the performance starts to diverge slightly from $C_{max}$. Due to hardware limitations and computational overhead in the pipeline, the curve starts to flatten at around $1\%$. In our experiments, values for $\tau$ reside in the range of $0.1$ to $0.01$, which equates to roughly 10$\times$ to 25$\times$ compression rates compared to the naive approach, which is almost ten times less efficient. 
\section{Limitations \& Future Directions}
FoVolNet works well on various data, as we have shown in the evaluation. It can benefit from more specialized training to address some of the edge cases we encountered during development, like high-frequency visual content or more pronounced transparency. Beyond this, there are several interesting extensions that we suggest here.

\vspace{0.03in}
\textbf{High-frequency Content.} Regions of a volume that contained high-frequency intensity shifts resulted in increased temporal flickering when using our network. In most cases, increasing the renderer's volume sampling rate would alleviate such issues. A more cost-effective approach is to purposely introduce such artifacts into the training data, which would reduce the overall severity of the issue. Another approach is to emphasize temporal loss terms by increasing their weight (sacrificing visual quality) or by introducing more adaptable terms like a GAN critique~\cite{Goodfellow2014}. 

\vspace{0.03in}
\textbf{Beyond Raymarching.} In this work, we showcase our technique on the example of a raymarching renderer. However, FoVolNet is easy to extend to Monte Carlo methods like volume path tracing. Here we see potential to stabilize framerates by cutting short long-running threads due to multiple bounces and reconstructing their results using constant-time neural networks. Support for other data types like particle volumes or flow fields could also be added. We encourage further research to explore the specifics of such extensions.


\vspace{0.03in}
\textbf{Neural Adaptive Sampling.} The approach presented here can be improved by predicting adaptive sample maps. Similar to Stengel et al.'s approach~\cite{Stengel2016}, both adaptive and foveated maps can be merged to maximize visual quality. Creating more off-focus sampling density could also improve the remaining issues with temporal stability. This extension increases sampling efficiency by replacing the naive uniform sampling in the periphery with an overall smarter approach.

\vspace{0.03in}
\textbf{Beyond the Screen.} High-fidelity immersive visualization of volume data is still out of reach today. However, with FoVolNet we achieve higher and more consistent framerates (Figure~\ref{fig:result_vortices_timings}). With further improvements to the network, this goal could be attained much sooner than with conventional rendering techniques. The inference overhead could be reduced to a point where real-time, high-fidelity rendering becomes possible. This would open up opportunities to utilize FoVolNet for immersive experiences of volume data in VR. 

\section{Conclusion}
We presented FoVolNet---a foveated neural reconstruction system for volume visualization. FoVolNet accelerates conventional volume rendering techniques by sparsely sampling the data and reconstructing the full-frame using deep learning. Our novel network design reconstructs the final rendering at high quality using a hybrid of direct and kernel prediction mechanisms. We show that FoVolNet is able to provide tremendous speed-ups at compression rates as high as 25$\times$ over the state-of-the-art control technique DeepFovea~\cite{KaplanyanDeepFovea} while preserving image quality close to the original. Our uncomplicated design makes it easy to be integrated into existing rendering pipelines.

It is our plan to combine this technique with neural representation compression techniques and streaming technology to push the field further towards real-time high-fidelity volume visualization on consumer hardware. There are numerous opportunities to use and extend this technique, and we hope to entice the visualization community to take up this pursuit.



\acknowledgments{This research is sponsored in part by the Department of Energy through grant SC-DE0019486 and Intel's oneAPI Centers of Excellence grant.}




\bibliographystyle{abbrv-doi-hyperref}

\bibliography{template}

\end{document}